\documentclass[twocolumn,showpacs,groupedaddress]{revtex4}
\usepackage{graphicx}
\usepackage{epsfig}
\usepackage{subfigure}
\usepackage{amsmath}

\begin{document}

\newcommand{\ket}[1]{|#1\rangle}
\newcommand{\bra}[1]{\langle#1|}
\newcommand{\Tr}{\text{Tr}}

\title{Effects of uncertainties and errors on Lyapunov control}
\author{X. X. Yi$^{1,2}$, B. Cui$^1$,
Chunfeng  Wu$^2$, and C. H. Oh$^2$}
\affiliation{$^1$School of Physics and Optoelectronic Technology\\
Dalian University of Technology, Dalian 116024 China\\
$^2$Centre for Quantum Technologies and Department of Physics,
National University of Singapore, 117543, Singapore }

\begin{abstract}
 Lyapunov control
(open-loop) is often confronted with uncertainties and errors in
practical applications. In this paper, we analyze the robustness of
Lyapunov control against the uncertainties and errors in  quantum
control  systems. The analysis is carried out through   examinations
of uncertainties and errors, calculations of  the control fidelity
under  influences  of the certainties and errors, as well as
discussions on the caused effects. Two examples, a closed control
system and an open control system, are presented to illustrate the
general formulism.
\end{abstract}

 \maketitle

\section{introduction}

Quantum control is the manipulation of the temporal evolution of a
system in order to obtain a desired target state or value of a
certain physical observable, realizing  it is a fundamental
challenge in many fields \cite{dong09,wiseman10,rabitz09}, including
atomic physics \cite{chu02}, molecular chemistry \cite{ rabitz00}
and quantum information \cite{nielsen00}. Several strategies of
quantum control have been introduced and developed from classical
control theory. For example,  optimal control theory has been used
to assist in control design for molecular systems and spin systems
\cite{khaneja01, alessandro01}. A learning control method has been
presented for guiding the control of chemical reactions
\cite{rabitz00}. Quantum feedback control approaches including
measurement-based feedback and coherent feedback have been used to
improve performance for several classes of tasks such as preparing
quantum states, quantum error correction and controlling quantum
entanglement \cite{wiseman93,doherty00}. Robust control tools have
been introduced to enhance the robustness of quantum feedback
networks and linear quantum stochastic systems
\cite{helon06,james08}.

Control systems are broadly classified as either closed-loop or
open-loop. An open-loop control system is controlled directly, and
only, by an input signal, whereas a closed-loop control system is
one in which an input forcing function is determined in part by the
system response. Among the open-loop controls, Lyapunov control has
been proven to be a sufficient control to be analyzed rigourously,
moreover, this control can be shown to be highly effective for
systems that satisfy certain sufficient conditions that roughly
speaking are equivalent to the controllability of the linearized
system.

Lyapunov control for quantum systems in fact use a feedback design
to construct an open-loop control. In other words, Lyapunov control
is used to first design a feedback law which is then used to find
the open-loop control by simulating the closed-loop system. Then the
control is applied to the quantum system in an open-loop way. From
the above description of Lyapunov control, we find that the Lyapunov
control includes two steps: (1) for any initial states and a system
Hamiltonian (assumed to be known exactly), design a control law,
i.e., calculate the control field by simulating the dynamics of the
closed-loop system, (2) apply the control law to the control system
in an open-loop way. Although some progress has been made, more
research effort is necessary in Lyapunov control, especially, the
robustness of quantum control systems has been recognized as a key
issue in developing practical quantum technology. In this paper, we
study the effect of uncertainties and errors on the performance of
Lyapunov control. The uncertainties come from initial states and
system Hamiltonian, and errors may occur in applying the control
field (control law). Through  this study, we show the robustness of
Lyapunov control against uncertainties  and errors. In particular,
the relation between the uncertainties and the fidelity is
established for a closed two-level control system and an open
four-level control system.

This paper is organized as follows. In Sec.\ref{problem}, we
introduce the Lyapunov control and formulate the problem. A general
formulism is given to examine the robustness of the Lyapunov
control. In Sec. \ref{exa}, we exemplify the general formulation in
Sec.\ref{problem} through a closed and an open quantum control
systems. Concluding remarks are given in Sec. \ref{con}.

\section{problem formulation}\label{problem}
A control quantum system can be modeled in different ways, either as
a closed system evolving unitarily governed  by a Hamiltonian, or as
an open system governed by a master equation. In this paper, we
restrict our discussion to a $N$-dimensional open quantum system,
and consider its dynamics as Markovian. The discussion is applicable
for closed systems, since closed system is a special case of open
system with zero decoherence rates.  Therefore we here consider a
system that obeys the Markovian master equation ($\hbar=1,$
throughout this paper),
\begin{eqnarray}
\dot{\rho}=-i[H,\rho]+{\mathcal L}(\rho) \label{mef}
\end{eqnarray}
with
 $${\mathcal
L}(\rho)=\frac 1 2\sum_{m=1}^M \lambda_m([J_m,\rho
J_m^{\dagger}]+[J_m\rho,J_m^{\dagger}]),$$ and
$$
H=H_0+\sum_{n=1}^F f_n(t)H_n,
$$
where $\lambda_m \ (m=1,2,...,M)$ are positive and time-independent
parameters, which characterize the decoherence and are called
decoherence rates. Furthermore,  $J_m \  (m=1,2,...,M)$ are  the
Lindblad operators, $H_0$ is the free Hamiltonian and $H_n \
(n=1,2,...,F)$ are control Hamiltonians, while $f_n(t) \
(n=1,2,...,F)$ are control fields. Equation (\ref{mef}) is of
Lindblad form, this means that the solution to Eq. (\ref{mef}) has
all the required properties of  physical density matrix at any
times. Since the free Hamiltonian can usually not be turned off, we
take nonstationary states $\rho_D(t)$ as target states that satisfy,
\begin{equation}
\dot{\rho}_D(t)=-i[H_0, \rho_D(t)].\label{rhod}
\end{equation}
The control fields $\{f_n(t), \ n=1,2,3,...\}$ can be established by
Lyapunov function. Define  $V(\rho_D, \rho)$,
\begin{equation}
V(\rho_D,\rho)=\mbox{Tr}(\rho_D^2)-\mbox{Tr}(\rho\rho_D),\label{lyaf}
\end{equation}
we find $V\geq 0$ and
\begin{equation}
\dot{V}=-\sum_n^F f_n(t)
\mbox{Tr}\{\rho_D[-iH_n,\rho]\}-\mbox{Tr}[\rho_D\mathcal{L}(\rho)].\label{dotv}
\end{equation}
For $V$ to be a Lyapunov function, it requires  $\dot{V}\leq 0$ and
$V\geq 0.$ If we choose a $n_0$ such that $f_{n_0}(t)
\mbox{Tr}\{\rho_D[-iH_{n_0},\rho]\}+\mbox{Tr}[\rho_D\mathcal{L}(\rho)]=0$,
and $f_n(t)=\mbox{Tr}\{\rho_D[-iH_n,\rho]\}$ for $n\neq n_0$, then
$\dot{V}\leq 0.$  With these choices, $V$ is a Lyapunov function.
Therefore,  the evolution of the open system with Lyapunov control
governed by the following nonlinear equations\cite{yi09}
\begin{eqnarray}
\dot{\rho}(t)&=&-i[H_0+\sum_{n} f_n(t)H_n,
\rho(t)]+\mathcal{L}(\rho),\nonumber\\
f_n(t)&=&\mbox{Tr}\{[-iH_n,\rho]\rho_D\}, \mbox{ for } n\neq
n_0,\nonumber\\
f_{n_0}(t)&=&-\frac{\mbox{Tr}[\rho_D\mathcal L
(\rho)]}{\mbox{Tr}\{\rho_D[\rho, iH_{n_0}]\}},\mbox{and} \nonumber\\
\dot{\rho}_D(t)&=&-i[H_0,\rho_D(t)] \label{nle1}
\end{eqnarray}
is stable in Lyapunov sense at least. In Eqs (\ref{rhod}) and
(\ref{lyaf}), we have identified $\rho_D(t)$ with target states,
this means that if a quantum system is driven into the target
states, it  will be maintained in these states under the action of
the free Hamiltonian. However, in practical applications, it is
inevitable that there exist errors and uncertainties in the free
Hamiltonian, in the initial states and in the control fields. These
uncertainties and errors would disturb the dynamics and steer the
system away from the target state. In the following, we suppose that
the uncertainties can be approximately described as perturbations
$\delta H_0$ in the free Hamiltonian, and as deviations $\delta
\rho_0$ in the initial state as well as fluctuations $\delta f_n \
(n $ may take $1,2,3,...)$ in the control fields. Then the actual
final state $\rho_R(t)$ of the control system starting from
$(\rho_0+\delta \rho_0)$ governed by Eq. (\ref{nle1}) with
$(H_0+\delta H_0)$ and $[f_n(t)+\delta f_n(t)]$ instead of $H_0$ and
$f_n(t)$ would be different from $\rho_D(t).$ We quantify the
difference between the target states $\rho_D(t)$ and the practical
states $\rho_R(t)$ by using the fidelity defined by $F(\rho_D,
\rho_R)=\Tr \sqrt{\rho_D^{\frac 1 2 }\rho_R \rho_D^{\frac 1 2 }}.$

For a Lyapunov control with negative gradient of Lyapunov function
in the neighborhood of target states, the controlled system state
will be attracted to and maintained in the target state, when there
are no uncertainties and errors. With uncertainties and errors, the
problem of robustness of the control system is not trivial, because
the Lyapunov-based feedback design for the control law would induce
nonlinearity in the control system. The LaSalle invariant
principle\cite{lasalle61} tells that the autonomous dynamical system
Eq.(\ref{nle1}) converges to an invariant set defined by
$\mathcal{E}=\{\rho_{in}: \dot{V}=0\}$, which is equivalent to
$f_n(t)=0, \ n=1,2,3,...$ by Eq.(\ref{nle1}). This set is in general
not empty and the final state will be in it. From Eqs. (\ref{dotv})
and (\ref{nle1}) we find that the invariant set is an
 intersection of all
sets $\mathcal{E}_n$ $(n=1,2,3,...,n\neq n_0)$, each  satisfies,
\begin{equation}
\mathcal{E}_n=\{\rho_{in,n}:
\Tr(\rho_dH_n\rho_{in,n}-H_n\rho_d\rho_{in,n})=0\}, \ \ n\neq n_0.
\end{equation}
Since the control fields are proportional to $\dot{V}$, the errors
in the control fields would change the invariant set. The
uncertainties in the initial state affect the invariant set in the
same way, and the uncertainties in the free Hamiltonian change the
target sets $\rho_d(t)$, leading to an invariant set different from
that without uncertainties. In the next section, we will illustrate
and exemplify the effect of errors and uncertainties on the fidelity
through simple examples.

\section{illustration}\label{exa}
In this section, we first introduce a Lyapunov control on a closed
two-level quantum system, then we study the robustness of this
Lyapunov control by examining the effects of uncertainties and
errors on the fidelity of control. Next, we extend this study into
open systems by considering a dissipative four-level system and
steering it to a target state in its decoherence-free subspace
(DFS).

We start with a closed two-level system described by the
Hamiltonian,
\begin{equation}
H=\frac\omega 2\sigma_z+f(t)\sigma_x\equiv H_0+H_1,
\end{equation}
where $H_0=\frac\omega 2\sigma_z$ denotes the free Hamiltonian of
the system, $H_1=f(t) \sigma_x$ is the control Hamiltonian with a
control field $f(t).$ We define one of the eigenstates of $H_0$, say
the ground state $|g\rangle$, as the target state, the Lyapunov
function in Eq. (\ref{lyaf})  for this closed system is then
\begin{equation}
V(|g\rangle, |\Phi(t)\rangle)=1-|\langle g|\Phi(t)\rangle|^2.
\end{equation}
For closed system, the  Liouvillian   ${\mathcal L}(\rho)$ vanishes,
thus we do not need to choose a control field $f_{n_0}(t)$ in Eq.
(\ref{nle1}) to cancel the drift term. The only control field $f(t)$
that can be derived from Eq. (\ref{nle1}) is,
\begin{equation}
f(t)=2{\mbox {Im}}(\langle
g|\sigma_x|\Phi(t)\rangle\langle\Phi(t)|g\rangle).
\end{equation}
Here $|\Phi(t)\rangle$ represents states at time $t$ starting from
an initial states $$|\Phi(0)\rangle =\cos \beta_0|e\rangle
+\sin\beta_0 e^{i\phi_0}|g\rangle,$$ under the action of the
Hamiltonian $H$ without any uncertainties and errors. We further
suppose that the uncertainties in the free Hamiltonian $H_0$ can be
described as a perturbation,
\begin{equation}
\delta H_0=\delta_x\sigma_x+\delta_z\sigma_z,
\end{equation}
and the uncertainties in the initial state $|\Phi(0)\rangle$ can be
characterized by replacing $\beta_0$ and $\phi_0$ with
$(\beta_0+\delta \beta_0)$ and $(\phi_0+\delta\phi_0)$,
respectively.
\begin{figure}
\includegraphics*[width=0.8\columnwidth,
height=0.6\columnwidth]{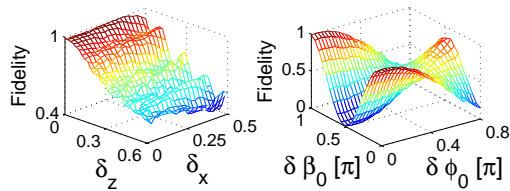} \vspace{-2cm} \caption{ (Color
online) Fidelity of Lyapunov control versus the uncertainties in the
free Hamiltonian (left) and in the initial states (right).
$\omega=4$ (in arbitrary units) and $\phi_0=\beta_0=\frac{\pi}{4}$
are chosen for this plot.}\label{f1}
\end{figure}
We describe the errors in the control fields $f(t)$ as fluctuations
$\delta(t)\cdot f(t)$ with random number $\delta(t)$.   With these
descriptions, the practical control system can be described by,
\begin{equation}
i\hbar \frac{\partial}{\partial t} |\psi(t)\rangle_R=[H_0+\delta
H_0+f(t)H_1(1+\delta(t))]|\psi(t)\rangle_R,\label{rc}
\end{equation}
with initial condition $ |\Phi(0)+\delta
\Phi(0)\rangle=\cos(\beta_0+\delta \beta_0)|e\rangle
+\sin(\beta_0+\delta \beta_0)e^{i(\phi_0+\delta\phi_0)}|g\rangle. $

\begin{figure}
\includegraphics*[width=0.8\columnwidth,
height=0.6\columnwidth]{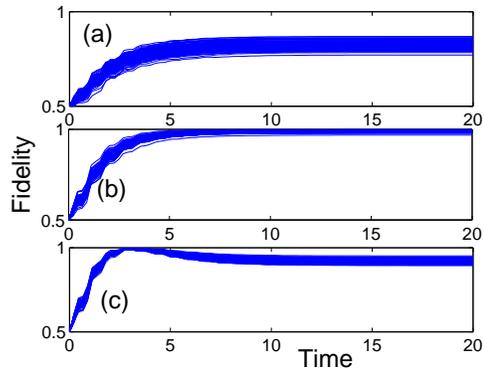}
 \caption{(Color online) Fidelity of Lyapunov control as a function of time.
 This plot shows the effects of fluctuations in the control field $f(t)$ on the fidelity.
(a), (b) and (c) are for different types of fluctuations. (a) The
fluctuation $\delta(t)$ was taken from $(-1)$ to zero; (b) from
$(-1)$ to (+1); and (c) from 0 to (+1). All fluctuations are taken
randomly. The other parameters chosen are the same as in Fig.
\ref{f1}. There are no uncertainties in the free Hamiltonian and in
the initial states. }\label{f2}
\end{figure}
We have performed numerical simulations for Eq. (\ref{rc}), selected
results are presented in figures \ref{f1} and \ref{f2}. Figure
\ref{f1} shows the control fidelity as a function of uncertainties
$(\delta_x, \delta_y)$ in the free Hamiltonian and uncertainties
$(\delta\beta_0,\delta \phi_0)$ in the initial state. Two
observations can be made from the figures. (1) The control fidelity
rapidly  depends on the uncertainties $\delta_z\sigma_z$, whereas it
is not sensitive to $\delta_x\sigma_x$, (2) the control fidelity is
an oscillating function of $\delta\beta_0$ and $\delta \phi_0$ with
different periods. These observations indicate that the Lyapunov
control on closed systems is robust against the uncertainties that
commute with the control Hamiltonian, while it is fragile with the
other uncertainties in the free Hamiltonian. This claim is confirmed
by Fig. \ref{f2}, where the effect of fluctuations in the control
field on the control fidelity is shown. One can clearly see from
figure \ref{f2} that there are almost no effects for the
fluctuations with zero mean on the fidelity. This can be understood
as follows. Since the fluctuations is randomly chosen for the
control fields, the net effect intrinsically equals to an average
over all fluctuations, which must be zero for fluctuation with zero
mean.

\begin{figure}
\includegraphics*[width=0.7\columnwidth,
height=0.5\columnwidth]{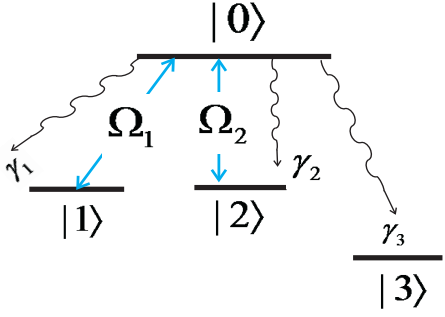}
 \caption{The schematic energy diagram. A four-level system with two degenerate stable states
 $|1\rangle$ and $|2\rangle$ in external laser fields.
 The two degenerate states are coupled to the excited state $|0\rangle$
 by two separate lasers with coupling constants $\Omega_1$ and $\Omega_2$, respectively. While the
 stable state $|3\rangle$ is isolated from the other levels. The excited state $|0\rangle$ decays to  $|j\rangle$
 ($j=1,2,3$) with decay rate $\gamma_j$.}\label{f3}
\end{figure}

Now we turn to another example that shows the robustness of Lyapunov
control on open quantum systems. We borrow the model in
Ref.\cite{yi10} shown in Fig.\ref{f3}, where a four-level system
coupling to two external lasers and being subject to decoherence has
been considered. The Hamiltonian of this system has the form,
\begin{equation}
H_0=\sum_{j=0}^2\Delta_j |j\rangle\langle j|+(\sum_{j=1}^2
\Omega_j|0\rangle\langle j|+h.c.),\label{freeH}
\end{equation}
where $\Omega_j \  \ (j=1,2)$ are coupling constants. Without loss
of generality, in the following the coupling constants are
parameterized as $\Omega_1=\Omega \cos \phi$ and
$\Omega_2=\Omega\sin\phi$  with
$\Omega=\sqrt{\Omega_1^2+\Omega_2^2}.$  The excited state
$|0\rangle$ is not stable, it decays to the three stable states with
rates $\gamma_1$, $\gamma_2$ and $\gamma_3$ respectively. We assume
this process is Markovian and can be described by the Liouvillian,
\begin{equation}
\mathcal L(\rho)=\sum_{j=1}^3\gamma_j(\sigma_j^-\rho\sigma_j^+-\frac
1 2 \sigma_j^+\sigma_j^-\rho-\frac 1 2
\rho\sigma_j^+\sigma_j^-)\label{diss}
\end{equation}
with $\sigma_j^-=|0\rangle\langle j|$ and
$\sigma_j^+=(\sigma_j^-)^{\dagger}.$ It is not difficult to find
that the two degenerate eigenstates
$|D_1\rangle=\cos\phi|2\rangle-\sin\phi|1\rangle,$
$|D_2\rangle=|3\rangle,$ of the free Hamiltonian $H_0$ form a DFS.
Now we show how to control the system to a desired target state
(e.g., $|D_1\rangle$) in the DFS. For this purpose, we choose the
control Hamiltonian
\begin{equation}
H_c=\sum_{j=1}^3 f_j(t)H_{j}\label{conH}
\end{equation}
 with
$ H_1$ is a 4 by 4 matrix with all elements equal to 1.
 $H_2=|D_1\rangle\langle D_2|+|D_2\rangle\langle
D_1|, H_3=|0\rangle\langle D_2|+|D_2\rangle\langle 0|. $ We shall
use Eq. (\ref{nle1}) to determine the control fields $\{f_n(t)\}$,
and choose
\begin{eqnarray}
|\Psi(0)\rangle&=&\sin\beta_1\cos\beta_3|0\rangle+\cos\beta_1\cos\beta_2|1\rangle\nonumber\\
&+& \cos\beta_1\sin\beta_2|2\rangle+\sin\beta_1\sin\beta_3|3\rangle
\label{inis}
\end{eqnarray}
as  initial states for the numerical simulation, where $\beta_1$,
$\beta_2$ and $\beta_3$ are allowed to change independently.  The
initial state written in Eq.(\ref{inis}) omits all (three) relative
phases between the  states $|0\rangle, |1\rangle, |2\rangle$ and
$|3\rangle$ in the superposition,  and satisfies the normalization
condition. $f_1(t)$ here is specified to cancel the drift term $\Tr
[{\mathcal L}(\rho)\hat{A}]$ in $\dot{V},$ this means that
$f_1(t)=-i\frac{\Tr({\mathcal
L}(\rho)\hat{A})}{\Tr([\hat{A},H_{1}]\rho)},$ $f_2(t)$ and $f_3(t)$
are determined by Eq.(\ref{nle1}).

\begin{figure}
\includegraphics*[width=0.8\columnwidth,
height=0.6\columnwidth]{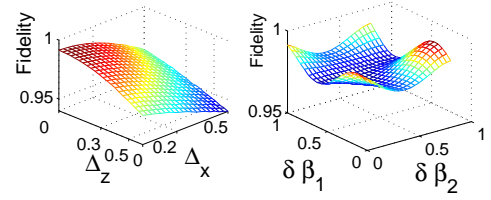} \vspace{-2cm}
 \caption{(Color online) The fidelity of control as a function of uncertainties in the free Hamiltonian (left)
 and in the initial state (right). The other parameters chosen are
 $\Omega=5, \phi=\frac{\pi}{5},
 \beta_3=\frac{\pi}{3}, \beta_1=\frac{\pi}{5}, \beta_2=\frac{\pi}{4},
 \gamma_1=\gamma_2=\gamma_3=\frac 1 3 \gamma,
 \kappa_2=1,$ $\Delta_0=4, \Delta_1=\Delta_2=2$
 and $
 \gamma=1.$ $\delta \beta_1$ and $\delta \beta_2$ are in unites of $\pi.$}\label{f4}
\end{figure}
\begin{figure}
\includegraphics*[width=0.8\columnwidth,
height=0.6\columnwidth]{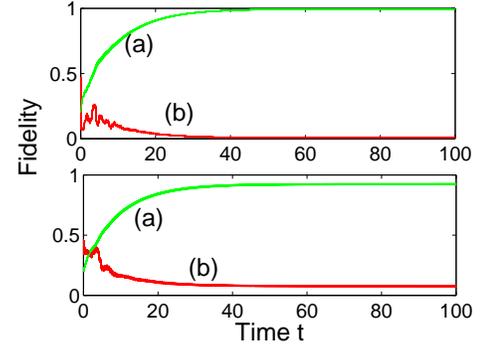}
 \caption{(Color online) Fidelity of Lyapunov control versus time $t$.
 The fluctuations $\delta_n \ (n=1,2,3,...)$ range
 from (-1) to (+1) for the upper panel, while from $(-1)$ to 0 (or 0 to (+1)) for the lower panel.
(a) in both panels denotes the probability in state $|D_1\rangle$,
while (b) in state $|D_2\rangle$.  The other parameters chose are
the same as in Fig. \ref{f4}. }\label{f5}
\end{figure}

We examine how the uncertainties in the free Hamiltonian and initial
states as well as the errors in the control fields $f_n(t),
(n=1,2,3,...)$ affect the fidelity of the control. These effects can
be illustrated by numerical simulations on
Eqs(\ref{freeH},\ref{diss},\ref{conH}), with the free Hamiltonian
$H_0$, the initial state $|\Phi(0)\rangle$ and the control fields
$f_n(t)$ replaced  by $(H_0+\delta H_0)$, $|\Phi(0)+\delta
\Phi(0)\rangle$ and $f_n(t)(1+\delta_n)$, respectively. Here,
\begin{eqnarray}
&&\delta H_0=\Delta_x(|0\rangle\langle 1|+|1\rangle\langle
0|)+\Delta_z(|0\rangle\langle 2|+|2\rangle\langle 0|),\nonumber\\
\ \ \nonumber\\
&&|\Phi(0)+\delta \Phi(0)\rangle=\sin(\beta_1+\delta
\beta_1)\cos\beta_3|0\rangle\nonumber\\
&&+\cos(\beta_1+\delta\beta_1)\cos(\beta_2+\delta\beta_2)|1\rangle\nonumber\\
&&+
\cos(\beta_1+\delta\beta_1)\sin(\beta_2+\delta\beta_2)|2\rangle\nonumber\\
&&+\sin(\beta_1+\delta\beta_1)\sin\beta_3|3\rangle,
\end{eqnarray}
and $\delta_n$ are random numbers ranging from $-1$ to $+1$. The
fidelity versus the uncertainties (characterized by $\Delta_x,
\Delta_z$, $\delta\beta_1, \delta\beta_2$) and errors ($\delta_n,
n=1,2,3,...)$ are presented in figures \ref{f4} and \ref{f5}.
Figures \ref{f4} and \ref{f5} tell us that the Lyapunov control on
open system with the target state $|D_1\rangle$ is robust against
the uncertainties in the initial state, and the fidelity is above
$95 \%$ when the uncertainties in the free Hamiltonian is bounded by
0.5 (in units of $\gamma$). The Lyapunov control is also robust
gainst the fluctuations in the control fields $f_n(t)$ as figure
\ref{f5} shows.
We note that the effects of fluctuations with zero mean are
different from that with non-zero mean. This can understood as an
average results taken over all fluctuations.

\section{concluding remarks}\label{con}
To summarize, we have examined the robustness of Lyapunov control in
quantum systems. The robustness is characterized by the fidelity of
the quantum state to the target state. Uncertainties in the free
Hamiltonian and in the initials states as well as the errors in the
control fields diminish the fidelity of control. The relation
between the uncertainties (errors) and the fidelity is established
for a closed two-level control system and an open four-level control
system. These results show that the Lyapunov control is robust
against the type of uncertainties which commute with the control
Hamiltonian, while it is fragile to the others. The fidelity is not
sensitive to zero mean random fluctuations (white noise) in the
control fields,  but it really decreases due to the non-zero
(positive or negative) mean  fluctuations.

\ \ \ \\
This work is supported by NSF of China under grant Nos 61078011 and
10935010, as well as the National Research Foundation and Ministry
of Education, Singapore under academic research grant No. WBS:
R-710-000-008-271.

\end{document}